\documentclass[12pt]{article}
\usepackage{epsfig}
\usepackage{amsmath,amscd,amsfonts,latexsym,amssymb,mathrsfs}
\newlength{\dinwidth}
\newlength{\dinmargin}
 \newtheorem{Definition}{Definition}[section]
 \newtheorem{Theorem}[Definition]{Theorem}

\newsymbol\rest 1316         

\def\cA{{\cal A}}
\def\cB{{\cal B}}

\def\cH{{\cal H}}

\def\cL{{\cal L}}
\def\cM{{\cal M}}

\def\cO{{\cal O}}

\def\cR{{\cal R}}

\def\bC{{\mathbb C}}

\def\bR{{\mathbb R}}
\def\RR{{\mathbb R}}

        \def\O{\Omega}

\newsymbol\bt 1202  

\newcommand{\hide}[1]{} 

\newcommand{\euL}{{\mathscr L}}

\begin{document}
\noindent
\begin{center}
{ \Large \bf Local covariant quantum field theory\\[6pt] over spectral geometries  }\\
\vspace*{1cm}
{\large {\sc Mario Paschke} and
{\sc Rainer Verch}}\\[10pt]
Max-Planck-Institut f\"ur Mathematik\\
in den Naturwissenschaften,\\
Inselstr.\ 22,\\
04103 Leipzig, Germany\\[6pt]
e-mail: paschke$@$mis.mpg.de, verch$@$mis.mpg.de
\end{center}
${}$\\[26pt]
{\small {\bf Abstract.} A framework which combines ideas from Connes' noncommutative geometry, or
spectral geometry, with recent ideas on generally covariant quantum field theory, is proposed in the
present work. A certain type of spectral geometries modelling (possibly noncommutative) globally hyperbolic
spacetimes is introduced in terms of so-called globally hyperbolic spectral triples.
The concept is further generalized to a category of globally hyperbolic spectral geometries whose morphisms
describe the generalization of isometric embeddings. Then a local generally covariant quantum
field theory is introduced as a covariant functor between such a category of globally hyperbolic
spectral geometries and the category of involutive algebras (or *-algebras). Thus, a local covariant
quantum field theory over spectral geometries assigns quantum fields not just to a single noncommutative
geometry (or noncommutative spacetime), but simultaneously to ``all'' spectral geometries, while
respecting the covariance principle demanding that quantum field theories over isomorphic spectral 
geometries should also be isomorphic. It is suggested that in a quantum theory of gravity a particular
class of globally hyperbolic spectral geometries is selected through a dynamical coupling 
of geometry and matter compatible with the covariance principle.}

\section{Introduction}
We know about space and time only from hearsay, the role of 
storytellers being played by the test particles with which we probe it. To be more to the point, our present picture
of the structure of spacetime at small scales solely relies on the fact that the experimental findings can all 
be described by quantum fields  which propagate on a fixed Lorentzian manifold. In particular, spacetime can locally be 
described as $\RR^4$ equipped with some Lorentzian metric. However, this description is only viable as long as we can 
neglect the dynamical coupling of the matter fields to spacetime at extreme energy scales.
 It has, in fact,  been argued (cf.\ e.g.\ \cite{DFR}) that
this dynamical coupling may lead  at sufficiently high energy densities to
some noncommutativity of spacetime, hence causing a fundamental obstruction to preparing states which are
localized in arbitrarily small regions. Even though somewhat speculative at the present stage, such arguments clearly
indicate that we shall have to reformulate quantum field theory in such a way that it allows to describe the
dynamical coupling of geometry and the quantized matter field.
This reformulation then requires the construction of a quantum theory which does not have to 
assume any fixed geometrical background,
but rather permits to reconstruct a (possibly noncommutative) spacetime from observed quantities, i.e. scattering data.

To undertake this task one might, in a first step, try to formulate QFT not over a single, fixed manifold, but over all 
noncommutative spacetimes simultaneously. After sufficient examination of the properties which such a 
theory inherits from the underlying set of geometries, one would try in a second step to turn the picture around,
and reconstruct spacetime from the states of a theory with these properties. In the final step, it is then attempted
to single out the theory which provides the physically correct coupling of geometry to matter.

As guiding principle for the first step of this program one should use the requirement of general covariance 
that underlies general relativity. ``QFT over all manifolds'' is thus characterized as a map which 
assigns to each (globally hyperbolic) manifold the algebra of observables of some QFT in a manner that is compatible
with the possibility to embed such manifolds into each other. This idea has been carried through in \cite{BFV} when restricted
to classical, i.e. commutative, spacetimes.

In this paper we briefly sketch  a proposal for a generalization of this approach to all noncommutative, globally hyperbolic
spin manifolds, which we shall define in a way that is similar to A. Connes' notion of noncommutative Riemannian Spin 
manifolds (cf.\ \cite{ConCMP182}).
As Connes' axioms are very restrictive, it is of course not clear a priori that they are fulfilled by every 
noncommutative space that may arise
dynamically via the coupling to quantum fields.  Thus the chosen class of noncommutative manifolds might still be
too small for our purpose.

However, in order to obtain a feasible, sufficiently constrained notion of QFT, we need ``enough'' geometrical structure
and in order to work out the correct meaning of ``enough'' in this context it seems to be a good advice not to proceed
in a haste, but to relax the axioms only step by step.

Connes' Noncommutative Geometry then constitutes a sensible vantage point as it not only contains
 a clear concept of general covariance, but it also
allows for a natural definition of a noncommutative gravitational field and its coupling to matter. Even more so, it describes
geometrical entities essentially in terms of scattering data. This might facilitate the formulation of 
quantum field theories over generic noncommutative spaces which have a well-defined operational meaning
as well as a clear physical interpretation. We hope that techniques from Noncommutative Geometry will also prove helpful
to describe the dynamical coupling of geometry to matter at the quantum level, i.e.\ to perform the final step of our program.
Since we define generally covariant QFT as a map from noncommutative spacetimes to algebras of observables, the second step
will ``simply'' consist in constructing the inverse map. 

It is far from clear whether this second step, not to speak of the third one, can be performed. However, being 
quite general in nature, our framework of ``spectral local QFT'' might well turn out helpful to reach much more modest goals.
First of all, it is presently completely unclear how to formulate QFT over {\em generic} noncommutative spaces. (Theories over
the Moyal-deformed $\RR^4$ have been investigated intensively over the last years.)
Since a meaningful formulation should not depend on the choice of ``coordinates'', i.e. generators of the algebra of functions
over the noncommutative spacetime, we believe that this aim can only be achieved in accordance with the principle of general
covariance.

Secondly,  in view of the model-independence of our framework  and as every theory of the gravitational field
should (sooner or later) be formulated in a background independent way, our proposed ``spectral local QFT''
might well serve as a common language
for the various different approaches to quantum gravity, thus facilitating to compare them among each other.

We therefore hope that our proposal is of interest to a wider readership, and for this reason we decided to 
provide the present, less technical but also less conclusive version of our approach.
 All the technical details of the definitions 
and proofs will be given in two forthcoming papers.

The organization of this article is as follows. In the second section we recapitulate the framework of local
generally covariant quantum field theories according to \cite{BFV}. In the third section we present some
basic material on Connes' spectral geometry and give the definition of the type of Lorentzian spectral
geometries we shall be considering (``globally hyperbolic spectral triples'', or ``ghysts''). Section 4 is
devoted to a generalization of these globally hyperbolic spectral triples and their covariance structures
in terms of a category, and a local covariant quantum field theory is introduced as a functor between
this category and the category of $*$-algebras. In section 5 we illustrate how this category of globally
hyperbolic spectral geometries is determined in the case that the spacetimes are all classical, using
however only the language of spectral geometry. We also indicate how to construct some simple
examples of covariant quantum field functors in the present setting, and the relation to Morita
equivalence of algebras. The paper is concluded by an outlook in Sec.\ 6.

\section{Local generally covariant quantum field theory}
\setcounter{equation}{0}
One of the critical elements of input in quantum field theory
resides in having to assume a fixed background spacetime, given once and 
for all. This is at variance with the concept of diffeomorphism covariance
which is central within general relativity, and moreover it incurres an undesirable global
feature into the formulation of the basic structure of quantum field theory even
at the level of observable quantities.

However, as has been pointed out in the recent publications \cite{VerSPST,HolWald1,BFV}, one can
eliminate this global feature and reach at a ``background-free'' setting for
quantum field theory by extending the framework of quantum field theory on
curved spacetime as we shall now summarize, following ref.\ \cite{BFV}. To this end, it turns out to be
helpful to make use of some very basic concepts of category theory;
see \cite{MacLane} as a general reference on these matters. We wish to
consider quantum fields on all ``reasonable'' (barring causal pathologies) spacetime
manifolds at once, and we view the collection of all those spacetime manifolds as a
category, denoted {\bf Man}. 

In greater detail, we define the object class ${\rm Obj}({\bf Man})$
of this category to coincide with the collection of all four-dimensional, globally hyperbolic
Lorentzian manifolds (each endowed with an orientation and time-orientation; additionally
one can assume that each spacetime carries a Lorentzian spin-structure, as will be done
below). We will denote the objects in ${\rm Obj}({\bf Man})$ by ${\cal M}$ for brevity, which
should more precisely be written
$({\cal M},g)$, where $g$ is the spacetime metric. Given two objects ${\cal M}_1$, ${\cal M}_2$
in ${\rm Obj}({\bf Man})$, we say that ${\cal M}_1 \overset{\psi}{\longrightarrow} {\cal M}_2$
is a morphism in the category {\bf Man} if $\psi$ is a $C^\infty$, isometric embedding of
${\cal M}_1$ into ${\cal M}_2$, preserving orientation and time-orientation, and such that
$\psi({\cal M}_1)$ is a globally hyperbolic submanifold of ${\cal M}_2$ in the following sense:
Given any two points $q,q'$ in $\psi({\cal M}_1)$, then each causal curve segment
in ${\cal M}_2$ having $q$ and
$q'$ as its endpoints is also contained in $\psi({\cal M}_1)$. The set of all morphisms 
${\cal M}_1 \overset{\psi}{\longrightarrow} {\cal M}_2$ will be denoted as ${\rm hom}({\cal M}_1,{\cal M}_2)$.
The composition of maps obviously serves as composition rule so as to combine
morphisms ${\cal M}_1 \overset{\psi}{\longrightarrow} {\cal M}_2$ and ${\cal M}_2 \overset{\psi'}{\longrightarrow} 
{\cal M}_3$
 into ${\cal M}_1 \overset{\psi' \circ \psi}{\longrightarrow} {\cal M}_3$.

Now the idea for a local generally covariant quantum field theory is basically to assign to
each globally hyperbolic spacetime, i.e.\ to each object of the category {\bf Man}, a quantum field 
theory in such a way that this assignment behaves covariantly. That is to say, a morphism 
${\cal M}_1 \overset{\psi}{\longrightarrow} {\cal M}_2$ should give rise to an identification of the
quantum field observables on ${\cal M}_1$ and on $\psi({\cal M}_2)$, where the latter are to
be viewed as sub-system of the quantum field observables assigned to ${\cal M}_2$. By ``identification''
we mean that the corresponding quantum field observables should have
an isomorphic algebraic structure. This can be expressed quite conveniently and efficiently by demanding
that a generally covariant quantum field theory be a covariant functor between the category {\bf Man}
and a second category, {\bf Alg}, which is a category of $*$-algebras. Thus, each object ${\cal A}$
in ${\rm Obj}({\bf Alg})$ is a (complex) $*$-algebra (simply an algebra over $\mathbb{C}$ with
a $*$-operation fulfilling $(AB)^* = B^*A^*$ and $(\lambda A + B)^* = \overline{\lambda} A^* + B^*$ for all
$\lambda \in \mathbb{C}$, $A,B \in {\cal A}$). Moreover, for ${\cal A}_1$, ${\cal A}_2$ being objects
in ${\rm Obj}({\bf Alg})$, ${\cal A}_1 \overset{\alpha}{\longrightarrow} {\cal A}_2$ is a morphism in
${\rm hom}({\cal A}_1,{\cal A}_2)$ if it is a faithful $*$-algebraic homomorphism (a linear map preserving the
algebra structures and $*$-operations). Therefore, one reaches at the following definition of
a local generally covariant quantum field theory:
\begin{Definition}
 {\bf \cite{BFV}} A {\bf local generally covariant quantum field theory} is a
covariant functor ${\sf A}$ from the category {\bf Man} of globally hyperbolic spacetime manifolds to the category 
{\bf Alg} of $*$-algebras, as represented by the following commuting diagram: 
$$
\begin{CD}
{\cal M}_1 @>\psi>> {\cal M}_2\\
@V{\sf A}VV     @VV{\sf A}V\\
{\sf A}({\cal M}_1)@>{{\sf A}(\psi)}>> {\sf A}({\cal M}_2)
\end{CD}
$$
\end{Definition}
Thus, a local generally covariant quantum field theory, symbolized by the functor {\sf A},
 assigns to each globally hyperbolic
spacetime manifold ${\cal M}$ a $*$-algebra ${\sf A}({\cal M})$, and to each isometric
embedding ${\cal M}_1 \overset{\psi}{\longrightarrow} {\cal M}_2$ of globally hyperbolic
spacetimes a $*$-algebraic embedding ${\sf A}({\cal M}_1) \overset{{\sf A}(\psi)}{\longrightarrow}{\sf A}({\cal M}_2)$,
so that the diagram above is commutative. It is more suggestive to write $\alpha_\psi$ in place
of ${\sf A}(\psi)$; with this notation, the covariance property of the functor {\sf A} expressed
by the commutativity of the diagram reads
$$ \alpha_{\psi'} \circ \alpha_{\psi} = \alpha_{\psi' \circ \psi} $$
for all morphisms ${\cal M}_1 \overset{\psi}{\longrightarrow} {\cal M}_2$ and ${\cal M}_2 \overset{\psi'}{\longrightarrow}
{\cal M}_3$.

Let us indicate some simple but instructive examples for local generally covariant quantum field theories.
Assume that to each globally hyperbolic spacetime ${\cal M}$ in the object class ${\rm Obj}({\bf Man})$ one
has assigned a hermitean 
scalar quantum field $\phi_{\cal M}$, i.e.\ for each ${\cal M}$, $\phi_{\cal M}: C_0^\infty({\cal M})
\to L(\mathscr{D}_{\cal M},\mathscr{H}_{\cal M})$ is an operator-valued distribution which maps, in
a linear and suitably continuous manner, any scalar test-function $f$ on ${\cal M}$ to a linear operator
$\phi_{\cal M}(f) :   \mathscr{D}_{\cal M} \to \mathscr{H}_{\cal M}$ where $\mathscr{H}_{\cal M}$ is
a Hilbert-space and $\mathscr{D}_{\cal M}$ a dense subspace, invariant under the action
of the field operators. Denoting by ${\sf A}({\cal M})$ the $*$-algebra generated by the field operators
$\phi_{\cal M}(f)$ as $f$ varies over $C_0^\infty({\cal M})$, we may envisage the situation where one obtains
an injective  $*$-algebraic homomorphism $\alpha_\psi: {\sf A}({\cal M}_1) \to {\sf A}({\cal M}_2)$ from
the definition
\begin{equation} \label{CovQF}
\alpha_{\psi}(\phi_{{\cal M}_1}(f)) = \phi_{{\cal M}_2}(\psi_*f) \,, \quad f \in C_0^\infty({\cal M}_1)\,,
\end{equation}
where $\psi_*f = f \circ \psi^{-1}$ is the push-forward
of $\psi \in {\rm hom}({\cal M}_1,{\cal M}_2)$. (This generalizes to quantum fields of
other tensor or spinor types.) In this case, one clearly finds that $\alpha_{\psi'} \circ \alpha_{\psi} =
\alpha_{\psi' \circ \psi}$ and therefore the just given definition of ${\sf A}({\cal M})$ together with
the identification
${\sf A}(\psi) = \alpha_\psi$ realizes a local, generally covariant quantum field theory. A more concrete example
is provided by the scalar Klein-Gordon field, which shows that the quantum field ``operators'' $\phi_{\cal M}(f)$
need not even be represented as operators on a Hilbert space. Namely, one can define (algebraically) 
the quantized scalar
Klein-Gordon field $\phi_{\cal M}$ on a globally hyperbolic spacetime $\cal M$ by requiring that
it assigns to each test-function $f$ on ${\cal M}$ a symbol $\phi_{\cal M}(f)$, linear in $f$, and subject
to the relations (i) $\phi_{\cal M}(f)^* = \phi_{\cal M}(\overline{f})$, (ii) $\phi_{\cal M}(\Box_{\cal M}f) = 0$,
(iii) $\phi_{\cal M}(f_1)\phi_{\cal M}(f_2) - \phi_{\cal M}(f_2)\phi_{\cal M}(f_1) = iE_{\cal M}(f_1,f_2)\cdot 1$,
where $\Box_{\cal M} = \nabla^a\nabla_a$ denotes the (massless) Klein-Gordon operator, or D'Alembert operator,
on ${\cal M}$, $E_{\cal M}$ denotes the causal propagator, or difference of advanced and retarded fundamental
solutions, for $\Box_{\cal M}$, and $1$ is a symbol for an algebraic unit element. The $\phi_{\cal M}(f)$,
together with these relations, generate a $*$-algebra which we will now denote by ${\sf A}({\cal M})$,
and from the unique existence of the fundamental solutions of the Klein-Gordon operator (see, e.g., \cite{DimockKG})
one can obtain quite easily that defining $\alpha_\psi$ as in \eqref{CovQF} induces an injective $*$-algebraic
homomorphism $\alpha_\psi: {\sf A}({\cal M}_1) \to {\sf A}({\cal M}_2)$.
We refer to \cite{BFV,HolWald1}
for further discussion of this example.
\\[6pt]
Let us continue with a couple of remarks.

First, one may consider other linear fields subject to hyperbolic field equations, like the Dirac field, the 
Proca field, and the Maxwell field, to obtain local general covariant quantum field theories by way of
similar constructions \cite{VerSPST,DimockD,DimockEM}. 

Secondly, one can specialize the framework and consider instead 
of the category ${\bf Alg}$ the category ${\bf CAlg}$ of $C^*$-algebras, wherein the objects are $C^*$-algebras,
and the morphisms are injective $C^*$-algebraic homomorphisms. While the functorial definition of a
local generally covariant QFT remains formally unchanged, dealing with $C^*$-algebras has some considerable 
advantages as far as mathematical and structural analysis is concerned. The Weyl-algebraic version of the
Klein-Gordon field \cite{DimockKG} provides the simplest example of a local generally covariant QFT
in the $C^*$-algebraic setting; see \cite{BFV} for details. Furthermore, when {\sf A} is a local generally
covariant QFT (defined with {\bf CAlg} in place of {\bf Alg}) then it turns out that the Haag-Kastler
approach to algebraic quantum field theory re-appears as a special case on each spacetime ${\cal M}$ in
${\rm Obj}({\bf Man})$: Defining for each open, globally hyperbolic subregion ${\cal O}$ of ${\cal M}$
the $C^*$-algebra ${\cal A}_{\cal M}({\cal O}) = {\sf A}(\iota_{{\cal M},{\cal O}}({\cal O}))$ where 
${\cal O} \overset{\iota_{{\cal M},{\cal O}}}{\longrightarrow} {\cal M}$ is the identical embedding map
viewed as a member of ${\rm hom}({\cal O},{\cal M})$, one clearly finds the condition of isotony,
${\cal A}_{\cal M}({\cal O}_1) \subset {\cal A}_{\cal M}({\cal O})$ for ${\cal O}_1 \subset {\cal O}$, fulfilled.
Moreover, the group of all orientation and time-orientation preserving, isometric diffeomorphisms
$\gamma : {\cal M} \to {\cal M}$ is represented by automorphisms $\alpha_{\gamma}$ on ${\cal A}_{\cal M}({\cal M})
= {\sf A}({\cal M})$ satisfying the more specialized condition of covariance, 
$\alpha_{\gamma}({\cal A}_{\cal M}({\cal O})) = {\cal A}_{\cal M}(\gamma({\cal O}))$. Again, we refer to
\cite{BFV} for a more extended discussion. In the case that all ${\cal A}_{\cal M}({\cal O})$ derived from
the functor ${\sf A}$ are causal in the sense that ${\cal A}_{\cal M}({\cal O})$ commutes 
elementwise with ${\cal A}_{\cal M}({\cal O}')$ whenever ${\cal O}$ and ${\cal O}'$ are
causally separated in ${\cal M}$, the local generally covariant QFT ${\sf A}$ will be called
{\sl causal}. This causality property is expected if the algebras ${\sf A}({\cal M})$ are formed by
observables; it is again fulfilled in the Klein-Gordon field example. 

As a third point, it is worth mentioning that the concept of local generally covariant quantum field theories
has recently led to very significant progress in QFT in curved spacetime. In particular,
time-ordered and normal-ordered products of the Klein-Gordon field could be constructed in such a
way as to yield local generally covariant fields in the sense that (an appropriate analogue of) eqn.\ \eqref{CovQF}
holds, and this has led to a considerable reduction of the renormalization ambiguity in the perturbative
construction of interacting quantum fields on generic curved spacetimes \cite{HolWald1,HolWald2,MorettiT}.
Moreover, the notion of local generally covariant QFT has pathed the way towards establishing strong
structural results in quantum field theory in curved spacetime, such as the connection between
spin and statistics and PCT \cite{VerSPST,HolPCT}.

\section{Noncommutative globally hyperbolic spacetimes}
There are many heuristic arguments for a noncommutative structure of spacetime at small distances, 
respectively sufficiently high energies, and we need not review them here. In any case, such 
speculations need experimental tests, and it is therefore desirable to 
extend the framework of quantum field theory to encompass theories over noncommutative spaces. Over the last years
(the renormalizability of) quantum field theories over the Moyal-deformed $\bR^4$ has been 
studied intensively. However, the restriction to such a particular
type of spacetime seems unsatisfactory as 
the information about the (noncommutative) nature of spacetime should be inferred from a measurement process rather than 
being assumed from the start. Thus, one would like to 
formulate quantum field theories over generic noncommutative spacetimes.
 Of course, this requires a feasible notion of noncommutative spacetimes, i.e. of
{\em (globally hyperbolic)  noncommutative Lorentzian manifolds}, that solve an appropriate generalization
of Einstein's equations. Unfortunately, this is still an open question in NCG, but there does 
exist an analogous notion of compact, Riemannian Spin manifolds, namely {\bf spectral triples} as defined by A.Connes in 
\cite{ConCMP182}; see also \cite{GB-F-V:book} for a comprehensive account.  \\
To become more concrete, a (real, even) spectral triple consists of five objects
$$(\cH, D, \cR,\gamma,J) , $$ fulfilling certain compatibilitity relations.
$\cR$ is a pre-$C^*$ algebra which, in the classical, i.e. commutative, case is taken to be the algebra of smooth functions
on some $d$-dimensional smooth and compact Spin-manifold $\Sigma$. Since $\Sigma$ is compact, the constant function
$f=1$ lies in $\cR$, i.e. $\cR$ is unital.\\
$\cR$ is represented on the Hilbert space $\cH$ as well as the unbounded operator $D=D^*$, the antilinear
isometry $J$ and $\gamma=\gamma^*$ with $\gamma^2=1$. \\
In the commutative case, $\cH$ can be interpreted as the space $L^2(\Sigma,S)$ of square integrable spinors
over $\Sigma$ with respect to a metric $h$ on $\Sigma$ that is uniquely specified by the operator $D$ and the algebra
$\cR$.    $D$ is then the Dirac-operator $\gamma^\mu \nabla_\mu$ on the spinor bundle over $\Sigma$ corresponding to $h$, while
$\gamma$ and $J$ can be viewed as the $d$-dimensional analogue of $\gamma_5$ and the charge conjugation, respectively.
In fact, the above mentioned compatibility conditions ensure the following. 
\begin{Theorem} {\bf \cite{ConCMP182}}
Let $\cR=C^\infty(\Sigma)$. Then to every spectral triple over $\cR$ there correponds uniquely
a metric $h$ and a spin structure $\sigma$ over $\Sigma$.
Vice versa, to each compact Riemannian Spin-manifold $(\Sigma,h,\sigma)$ one can 
(almost uniquely) associate a spectral triple over $C^\infty(\Sigma)$.
\end{Theorem}
Thus, for commutative algebras $\cR$ there is (up to an irrelevant freedom in the construction of $D$) a 
one-to-one correspondance between spectral triples and compact Riemannian Spin-manifolds.
There do, however, also exist many examples of spectral triples for noncommutative algebras $\cR$, e.g.
the noncommutative torus (cf.\ \cite{GB-F-V:book} and refs.\ cited there),
where the algebra $\cR$ is generated by two unitaries $U,V$ such that
 \[ UV = e^{i\theta}VU, \qquad\qquad \theta\in \RR .\]
In that sense, spectral triples can be viewed as a natural 
extension of the category of compact Riemannian manifolds.

Moreover, for all spectral triples it is possible to define not only an (unquantized)
fermionic action $S_F(\psi)= \langle \psi, D\psi \rangle$,  $\psi\in \cH$, but also the analogue of the Einstein-Hilbert-action
by setting $S_{Grav}(h) = S_{Grav}(D) = Tr(\chi(D))$ where $\chi$ is some appropriate function, 
chosen in particular such that 
$\chi(D)$ is trace-class. One might therefore also define generalized Einstein manifolds as
extrema of the action functional. We should stress that in the philosophy of Connes' Noncommutative Geometry it is 
not desired to consider other matter fields apart from the elementary fermions (leptons and quarks) and the gravitational
field. The strong and electroweak gauge bosons, but also the Higgs field, are considered to be 
signals of a noncommutativity of spacetime at small scales. More precisely, they will
be a part of the ``gravitational field'' $D$ for suitable noncommutative spaces. 
In fact (cf. \cite{ConCMP182}) the above spectral action reproduces the full (classical) Lagrangian of the standard model
of elementary particle physics if one considers  a certain spectral triple for the algebra $\cR = C^\infty(\cM)\otimes \cA_F$,
where $$\cA_F = \bC \oplus M_3(\bC)\oplus \mathbb{H}. $$ Here  $\mathbb{H} $ denotes the algebra of quaternions.

However, as mentioned above, up to now there does not exist a completely satisfactory generalization of spectral triples
to (noncompact) Lorentzian manifolds. The main difficulties stem from the fact that generic Lorentzian manifolds are noncompact,
i.e. that the corresponding algebras $\cR$ do not possess a unit element, leading to many technical complications.
The most severe problem raised by this fact, however, is that
for the reconstruction of the various vector bundles over $\cM$ that are of central importance in the definition of
spectral triples, one has to ``fix boundary conditions at infinity''. In algebraic terms, one
has to adjoin a unit to $\cA$, which can be done in many inequivalent ways. In the commutative case,
the appropriate way to do so in each specific case is well known, but a feasible generalization to noncommutative algebras
has not yet been found. 
Note also  that it is not possible to define the spectral action  $S(D) = Tr(\chi(D))$ , i.e. the noncommutative
generalization of the Einstein-Hilbert action for a Lorentzian signature of the metric: 
Being a hyperbolic rather than an elliptic operator,
the eigenvalues of the Dirac-operator $D$  are then infinitely degenerated, and hence
the operator $\chi(D)$ cannot be trace class.  
 
We should remark that in \cite{Stroh} a description of
{\em compact} Lorentzian manifolds that is very close to the notion of spectral triples has been
proposed. (See also \cite{Mor} for a different approach towards non-compact Lorentzian geometries.)
 But for our purpose of extending the framework of generally covariant quantum field theory to noncommutative
spacetimes we shall need a notion of noncommutative globally hyperbolic spacetimes, which are not compact.\\

The pragmatic, yet physically motivated approach 
followed here has been initiated in
\cite{Hawks} and has been further developed in \cite{KP}.

The basic idea is to describe a globally hyperbolic Lorentzian spacetime $\cM$
in terms of a Time-Space splitting (as in the Hamiltonian approach):
$$ \cM = \mathbb{R} \times \Sigma $$
and to model the ``Space'', i.e. the leaves $\Sigma_t = \{t\} \times \Sigma$ of the foliation
  for all $t\in \RR$  as a spectral triple.
This is, of course, only possible if $\Sigma$ is compact.

Given the metric information on each $\Sigma_t$,
in order to reconstruct the full geometric information about the 
foliation, one only needs the lapse function $N$ and shift vector ${\bf N}$, defined
by $$ \frac{\partial}{\partial t} = N\partial_0 + {\bf N}^i\frac{\partial}{\partial x^i}, $$ 
where $\partial_0$ denotes the unit vector that is orthogonal to $\Sigma$ in $\cM$ and
$t$ is the foliation-time coordinate.

As it turns out, lapse and shift can be reconstructed from the
Dirac-Hamiltonians $\{H_{t}\}$ at each time $t$ and
 $\beta_t$, i.e. the analogue of $\gamma_0$ \cite{KP}.
Indeed, writing the full Dirac-Operator $D$ on $\cM$ schematically as
\[ D = i\beta_t \partial_0 + D_\Sigma \] where $D_\Sigma$ denotes the
(elliptic) spatial Dirac-operator on $\Sigma_t$ (up to a zeroth order term coming from the spin connection,
which is not relevant in what follows), one obtains
$$\beta_t H_t = N D_\Sigma - i\beta_t {\bf N}^k\frac{\partial}{\partial x^k} ,$$ while $\beta_t^2 = N^2$.
Note that, up to a term of zeroth order, $D_\Sigma$ can be reconstructed as
$$ D_\Sigma \sim [H_t,\beta_t].$$
Moreover, it turns out to be much more convenient to work with the Dirac-Hamiltonian
$H_t$ rather than the spatial Dirac-operator $D_\Sigma$ for an algebraic characterization of globally hyperbolic manifolds,
as the unitary time evolution operators $U_{t,s}$ generated by the family $\{H_{t}\}$
can be used efficiently to describe the (smooth) ``glueing together'' of the spatial
hypersurfaces $\Sigma_t$:

As E. Hawkins has pointed out \cite{Hawks}, the  Cauchy-surfaces $\Sigma_t$ for all $t$ can all be described
using the same Hilbert space $\cH$, namely
the space of (spatially square integrable) {\em solutions of the Dirac-equation on $\cM$}.  
On this Hilbert space $\cH$ one then represents the families $H_t$, $\beta_t$ and $\gamma_t$ as well as the
algebras $\cR_t$ of functions on $\Sigma_t$. The charge conjugation $J$ is time-independent.\\
With $\cH$ being the space of solutions to $H\psi = i\frac{\partial}{\partial t}\psi$
it is then required that the unitary time evolution operators $U_{t,s}$ generated by $\{H_{\tau}\}$
intertwine $\cR_s$ and $\cR_t$ for all fixed $t,s$. Note that the family $\cR_t$ can be viewed
as a smooth bundle of algebras over the real line. We can thus also define the algebra
$\cL$ of smooth sections (vanishing at $t\to \pm \infty$) of this bundle. The elements
of $\cL$ can then be identified with functions over $\cM = \RR \times \Sigma$.

With these notations we can now define the analogue of spectral triples for (spatially compact)
globally hyperbolic spacetimes:

\begin{Definition}   A {\bf globally hyperbolic spectral triple (ghyst) } of dim.\ $1 + n$
is an ordered collection of objects
$$ \boldsymbol{L} = (\cL,\mathcal{H}^\cL,D^\cL,J^\cL,\gamma^\cL)\footnote{we have added a superscript $\cL$ to distinguish the data 
of the Lorentzian case from those of spectral triples}$$
where:
\begin{itemize}
\item $\mathcal{H}^\cL = L^2(\mathbb{R},\mathcal{H})$,\quad  
$\mathcal{H}^{\cL\,\infty} = C^\infty_0(\mathbb{R},\mathcal{H}^\infty)$\\
\item $(D^\cL\psi)(t) = i\beta_t\frac{\partial}{\partial t}\psi(t) + \beta_t H_t\psi(t)$, \quad 
   $\psi(\,.\,) \in \mathcal{H}^{L\,\infty}$ \\
\item The elements of $\cL$ act by ``multiplication'': \\
                $(a\psi)(t) = a(t)\psi(t)$ \\
\item $(J^\cL\psi)(t) = J\psi(t)$, \quad $(\gamma^\cL\psi)(t) = \beta_t\gamma_t\psi(t)$
\end{itemize}
These objects are subject to certain compatibility conditions. In particular, it is required that the data
$(\cR_t,\cH, D_\Sigma = [H_t,\beta_t],\gamma_t, J)$ fullfill for all $t$ the axioms of spectral triples, thus describing
the geometry of $\Sigma_t$.
\end{Definition}

Needless to say, the mentioned compatibility conditions ensure the
\begin{Theorem}
Let $\cL= C^\infty_0(\cM)$.
To each ghyst $\boldsymbol{L}$ over $\cL$ there exists a unique spatially compact globally hyperbolic Lorentzian  Spin manifold
$(\cM, g, \sigma)$. Conversely, to each such manifold
one can construct a ghyst.
\end{Theorem}
The proof will be given in a forthcoming paper.

Since the definition of a ghyst works with a fixed foliation, the construction of ghysts for a given 
globally hyperbolic Spin-manifold  is not as unique as the construction of spectral triples for Riemannian manifolds.
This then raises the problem to identify ghysts, $\boldsymbol{L_1}$ and  $\boldsymbol{L_2}$ say, 
that describe the same spacetime $(\cM, g, \sigma)$. Note that this identification is also essential for the
formulation of generally covariant quantum field theories over ghysts.
Fortunately the answer to this question is surprisingly simple:

Evidently,  $\boldsymbol{L_1}$ and  $\boldsymbol{L_2}$ will describe the same geometry
if they are unitarily equivalent, i.e.\ if there  
exists a unitary $U\, :\,\mathcal{H}^\cL_1\to \mathcal{H}^\cL_2 $, intertwining all the objects
of the two ghysts:
$$  U X_1 U^*= X_2 \,, \qquad \quad X = \cL,\, J^\cL,\, \gamma^\cL \, \beta \,,$$
apart from the Dirac-Operators $D_1, D_2$, where instead it is required that
$$ U ([D^\cL_1,a_1]) U^* = [D^\cL_2,U a_1 U^*]\,, \qquad\qquad  a_1\in  \cL_1 \, . $$
(The metric is reconstructed from the commutators $[D^L_1,a_1]$, therefore this condition on $D$ is sufficient.
It would not always be possible to unitarily intertwine the Dirac-Operators corresponding to different
foliations of the same manifold.)
As it turns out, the converse statement is also true (and will be proven elsewhere):
\\[6pt] 
{\em
Two globally hyperbolic spectral triples describe the same spacetime if and only if they are unitarily equivalent
in the above sense.}
\\[6pt]
In particular, the notion of ghysts together with this equivalence then provides a complete characterization
of (spatially compact) globally hyperbolic manifolds, i.e.\ of the commutative case. \\
Even more so, each noncommutative spectral triple can be used to construct a (family of) ghyst(s), thus establishing
the existence of nontrivial noncommutative examples of globally hyperbolic manifolds.

A major drawback of the notion of a ghyst is certainly that it is (presently) restricted to spacetimes with
a commuting time variable, or, to be more precise,
to spacetimes where the operator $\frac{\partial}{\partial t}$ exists as a derivation on $\cL$.  
There are, however, heuristic arguments to the effect that one should not expect such a time variable to exist
at high energies, see e.g. \cite{DFR}. It would then be desirable to formulate also models with a noncommutative time
that can be tested experimentally.

Nevertheless, we expect that it is not  possible to formulate consistent quantum field theories
over spacetimes with an arbitrarily noncommutative time, but that
the requirement of a meaningful physical interpretation will lead to restrictions on the commutator of time and
space coordinates. 
One might, for instance, require that at sufficiently low energies a ``classical'' time variable
emerges.

It might therefore be the better strategy to first formulate quantum field theories over 
spacetimes with a commuting time and then to relax this condition in a next step. Proceeding in this way
one might hope to gain better insight into the physical interpretation of
quantum theories on noncommutative spaces, and more importantly the specific 
form of noncommutativity that is expected to arise via quantum effects of the gravitational field.

Before formulating general covariant quantum field theories over ghysts, we should also mention a 
major advantage of the notion of ghyst: It is essentially based on data like the (spectrum of the) Hamilton-operator,
the grading, charge conjugation, i.e. on data which not only have a clear operational meaning, but 
even more so have a direct physical interpretation and are, in principle, accessible in scattering experiments.

\section{Local general covariant QFT over globally hyper\-bo\-lic spectral triples}

Now we wish to extend the approach to generally covariant QFT
outlined in Section 2 upon replacing (the categories of) globally hyperbolic spacetimes by
globally hyperbolic spectral triples. 

Already at the very beginning one encounters the difficulty that the class of classical, i.e. commutative,
ghysts is smaller than Obj(${\bf Man}$) for the reason that a classical ghyst corresponds to a spatially compact
globally hyperbolic spacetime. On the other hand, the possibility to embed
globally hyperbolic spacetimes isometrically into each other was crucial for the morphism structure
of the category ${\bf Man}$, and even more so for the definition of a local generally
covariant QFT as a covariant functor. Yet there exist only very specific non-trivial isometric embeddings
of a given spatially compact globally hyperbolic spacetime into a second one: Such embeddings can only
be proper in the ``temporal'' direction, but not in the ``spatial'' direction.

Without sufficiently many isometric embeddings, however, a functorial definition of a generally covariant
quantum field theory along the lines of Section 2 would appear much less powerful.
What is needed, therefore, is the concept of a ghyst which, in the classical case, would correspond to a globally
hyperbolic spacetime which is spatially non-compact. From the way the ghysts have been defined, this would essentially 
amount to a concept of Riemannian spectral triples generalizing also non-compact Riemannian manifolds, 
and as we have mentioned, this problem
has up to now not reached at a fully satisfactory solution.
At any rate, luckily, we can do with less, since we can leave the precise definition of ``spatially non-compact'' 
ghysts to some extent unspecified. It is sufficient for our purpose to define a notion of ``spatially non-compact ghysts
embedded into proper (spatially compact) ghysts'', and the morphisms between them.\\
Here again, basic concepts of category theory
prove useful, and to begin we gather all the (proper) ghysts in the object class Obj(${\bf ghyst}$)
of a category ${\bf ghyst}$. Given two ghysts $\boldsymbol{L_1}$ and  $\boldsymbol{L_2}$, we say that $u$
is in ${\rm hom}(\boldsymbol{L_1},\boldsymbol{L_2})$ if $\boldsymbol{L_1}$ and $\boldsymbol{L_2}$ are unitarily equivalent
by an unitary intertwiner $U\;:\; \cH_1^\cL \to \cH_2^\cL $ and $u={\rm Ad}U$. (Thus, ${\bf ghyst}$ is an
isomorphism category.)

We shall now enlarge the category ${\bf ghyst}$. To this end, we say that a triple $({\bf \euL}, V,\boldsymbol{L}) $
is an object in the category ${\bf Ghyst}$ if
\begin{enumerate}
\item $\euL = (\ell, \cH^\ell,D^\ell, J^\ell,\gamma^\ell)$ is a collection of objects 
     where $\cH^\ell $ is a Hilbert space operated upon by the algebra $\ell$ ;
      $D^\ell $ is an (essentially selfadjoint) operator on a dense domain 
      $\left(\cH^\ell\right)^\infty $ (which is left invariant under the action
      of $\ell, J^\ell $ and $\gamma^\ell$.)
\item $\boldsymbol{L}$ is an object in Obj(${\bf ghyst}$), and $ V\;:\;\cH^\ell \to \cH^\cL $ is an isometry so that
      \begin{eqnarray} \ell & = & V^*\cL V , \nonumber\\ 
       X^\ell & = & V^* X^\cL V \qquad \qquad\quad {\rm for} \quad X= J,\gamma ,  \label{cova} \\
      {\rm and} \qquad \qquad  [D^\ell, \cdot\,] & = & {\rm Ad}\,V^* \circ [D^\cL, {\rm Ad}\,V(\cdot)\,] \nonumber
     \end{eqnarray}
\item The data $\ell,D^\ell, J^\ell$ and $\gamma^\ell$ satisfy the same algebraic relations amongst 
      each other as $\cL, D^\cL,J^\cL, \gamma^\cL$.
\end{enumerate} 
Thus the ``source entries'' $\euL$ of the triple $(\euL,V,\boldsymbol{L}) $ are entities which 
are embedded by an isometry into a proper ghyst $\boldsymbol{L}$ and have themselves almost the full structure of a ghyst.
To see that $\euL$ is embedded into $\boldsymbol{L}$ (actually, into a part of $\boldsymbol{L}$) by the isometry
$V$, note that the first condition of \eqref{cova} is equivalent to $V \ell V^* = P \cL P$ where $P = VV^*$ is
the projector onto the range of $V$. The remaining conditions have a similar equivalent counterpart involving
${\rm Ad}\,V$ on the left hand side.

However, some properties of a ``proper'' ghyst are lacking on the side of $\euL$.
E.g., in the classical case, $\euL$ could represent a spatially open globally hyperbolic spacetime (with spin structure),
which is embedded into a spatially compact globally hyperbolic spacetime (with spin structure) represented by
$\boldsymbol{L}$, thus representing a globally hyperbolic submanifold
of $\boldsymbol{L}$. This possibility is gained by allowing $V$ to be an isometry instead of a unitary. Thus, one can think
of $\euL$ as being foliated into ``almost'' spectral triples $(\cR_t, \cH_t,D_\Sigma, \gamma_t, J)$. 
The isometric embedding of $\euL $ into $\boldsymbol{L}$
nevertheless ensures that there is an ambient algebraic structure sufficiently constrained so that
-- thinking of the classical case -- all geometric information on $\euL$ can be retrieved via the embedding.

It is useful to visualize the triple $(\euL,V,\boldsymbol{L}) $ as an arrow 
$ \euL\stackrel{v}{\longrightarrow}\boldsymbol{L}$,
where $v = {\rm Ad}V$. Thus, ${\bf Ghyst}$ is formally a category of (horizontal) arrows as objects.
Then given two such objects $\euL_1 \stackrel{v_1}{\longrightarrow}\boldsymbol{L_1}$ (or simply $v_1$)
and $\euL_2 \stackrel{v_2}{\longrightarrow}\boldsymbol{L_2}$ (resp., $v_2$) 
in Obj(${\bf Ghyst}$) we say that a pair $(w^\ell,w^\cL)$ is a morphism (or: vertical arrow) 
in hom$(v_1,v_2)$ if $w^\ell ={\rm Ad}\,W^\ell$,  $w^\cL ={\rm Ad}\,W^L$
with an isometry $W^\ell \; :\; \cH_1^\ell \to \cH_2^\ell $ and a partial isometry
$W^L\;:\; \cH_1^\cL \to \cH_2^\cL$ so that the following diagram
is a commuting square: 

\begin{equation*}
\begin{CD}
\euL_1 @>v_1 >> \boldsymbol{L_1}\\
@ V{w^\ell}VV     @ VV{w^\cL}V\\
\euL_2@>v_2>> \boldsymbol{L_2}
\end{CD}
\end{equation*}

The intertwining relations along the arrows are to be understood in a manner analogous to (\ref{cova}). 
Commutativity of the diagram means, in particular,
\[w^\cL\circ v_1 = v_2\circ w^\ell . \] 
We shall write ${\bf w}$ for the pair $(w^\ell,w^\cL)$. The composition
${\bf w_{32}}\circ {\bf w_{21}}$ for ${\bf w_{21}} \in {\rm hom}(v_1,v_2)$ and ${\bf w_{32}} \in {\rm hom}(v_2,v_3)$
is defined component-wise: 
$${\bf w_{32}}\circ {\bf w_{21}} = (w^\ell_{32}\circ w^\ell_{21}\,,\,w^\cL_{32}\circ w^\cL_{21}).$$
It should be clear that this composition rule is associative and that there are unit isomorphisms in each hom$(v,v)$.
Thus we have completed the definition of the category ${\bf Ghyst}$.

However, now the category ${\bf Ghyst}$ may be too large for the purposes of formulating generally covariant 
quantum field theory in a functorial fashion. Indeed, thinking of the classical situation, Obj$({\bf Ghyst})$
could contain also open Lorentzian submanifolds of some spatially compact globally hyperbolic spacetime
which aren't {\em globally hyperbolic} submanifolds. Thus, one would like to restrict this freedom
a bit and this can be done by selecting  sub-categories of   ${\bf Ghyst}$.
The initial datum for a generally covariant quantum field theory over spectral geometries is therefore
a  sub-category of the category  ${\bf Ghyst}$.

\begin{Definition}
A {\bf generally covariant quantum field theory over spectral geometries}, or spectral local QFT, consists of a choice
of a  sub-category ${\bf Geo}$ of ${\bf Ghyst}$ and a functor ${\sf A}$ from ${\bf Geo}$ to ${\bf Alg}$:
\begin{equation*}
\begin{CD}
 v_1 @>{\bf w} >> v_2\\
@V{{\sf A}}VV     @VV{{\sf A}}V\\
{\sf A}(v_1)@>{{\sf A}({\bf w})}>> {\sf A}(v_2)
\end{CD}
\end{equation*}
\end{Definition}

It is important to note that we have not specified how the sub-category ${\bf Geo}$ is to be chosen.
In the classical (commutative) case it is possible to fully characterize ${\bf Geo}$, as we shall
illustrate in the next section. However, it is at this point not clear how to do so over
noncommutative spaces, and this might seem disturbing at first sight.

On the other hand, as we are mainly concerned with noncommutative spacetimes because they might play
an essential  role in a quantum theory of the gravitational field and its coupling to matter,
one would rather like to gather information about the correct choice of ``open submanifolds''
by examining the dynamical coupling of geometry and matter at the quantum level.
Thus, in the light of this philosophy, it is quite gratifying that for 
the definition of generally covariant  quantum field theories we need not fix ${\bf Geo}$ completely.   

In fact, for the definition and the basic properties of the QFT-functor only its covariance is
needed, and this covariance is already a very restrictive requirement.
It is therefore also sufficient that the category ${\bf Ghyst}$ and hence its subcategory ${\bf Geo}$
are only defined ``relatively'', i.e. via their embeddings.  
In view of this strategy it might actually be  more general and natural to
replace also the target of the quantum field functor by arrows, i.e. by embeddings of
algebras into each other.

Beyond doubt, it would be even more general to formulate generally covariant quantum field theories in terms of
(the convex set of) physical states rather than algebras of observables. The emergence of a spacetime that can then be 
reconstructed via the observables is certainly a property of some subset of states of the full quantum theory.
Moreover, in many cases of physical relevance, like the appearance of superselection sectors (``charge sectors''),
it is more appropriate to work with states (and charged fields interpolating between the different sectors of states)
rather than with the algebra of observables (which have to commute with the charges by definition).
However, on the technical side, it is much more involved to formulate the requirement of general covariance
upon replacing the algebra of observables by the states (see \cite{BFV} for a detailed discussion).
It is also still unclear how to reformulate additional requirements on the functor ${\sf A}$, for instance
locality, in this case -- we should refer to \cite{BS2} for a promising recent attempt to deduce
locality from properties of the physical states, however.

\section{An illustration and an example}

\subsection{The construction of the subcategory ${\bf Geo}$}

We would now like to illustrate the construction of {\em globally hyperbolic} submanifolds and
in particular the isometrie $V$ in the commutative case. For simplicity and definiteness we shall
confine ourselves to $1+1$ dimensions. However, all the subsequent constructions immediately extend to
the higher dimensional case. 

Since ghysts are defined via a foliation of the globally hyperbolic Spin manifold $(\cM, g, \sigma)$, where 
$\cM=\RR\times \Sigma$, into spectral triples over the algebra $C^\infty(\Sigma)$,
the starting point is to first examine the construction of the analogous isometries for spectral triples.

Suppose we are given a spectral triple 
$(\cR,\cH,D,J,\gamma)$ over $\cR=C^\infty(\Sigma)$ and a smooth submanifold $\O \subset\Sigma$ having the same
dimension as $\Sigma$.
Let $\cH^\infty = \left(L^2(\Sigma,S)\right)^\infty$ denote the subspace of $\cH$ formed by the {\em smooth} 
square integrable sections of the spinor-bundle
over $\Sigma$. 

Consider now the space $\cH^\infty_\O$ of elements of $\cH^\infty$ with support in $\O$.
Note that, because they are required to be smooth, the elements of $\cH^\infty_\O$ vanish 
together with all derivatives at the boundary $\partial \O$. Then define
$E_\O \in\cB(\cH)$  as the orthogonal projector from $\cH$ onto the closure of $\cH^\infty_\O$.
Evidently, $\cR_\O := E_\O \cR E_\O$ is then just the algebra of all smooth functions on $\O$
which can be extended to smooth functions on $\Sigma$. 

Likewise, $D_\O = E_\O D E_\O$ will be the Dirac-operator on $\O\backslash\partial\O$ and similarly for
$J$ and $\gamma$. Note that we need to work with smooth sections to ensure that $D_\O$ is a differential operator
of first order. As we have seen, this results in the loss of all information about the boundary of $\O$.
Thus, with our definition, we can only describe submanifolds of spectral triples without a boundary.
That does, however, not cause any problems for our purpose, as we are actually interested
in constructing {\em open} submanifolds of $\Sigma$ anyway.

In $1+1$ dimensions
$\Sigma$ is homeomorphic to a circle $S^1$ and we shall now construct globally hyperbolic submanifolds
of $(\cM=\RR\times S^1, g,\sigma) $, so-called diamonds. The pregiven ghyst over $\cM$ will be denoted by
$\boldsymbol{L} = (\cL,\cH^\cL,D^\cL,J^\cL,\gamma^\cL)$. 

Of course, taking any smooth open submanifold $\cO \subset \cM$ and constructing a projector $P_\cO$ in 
a fashion analogous to 
$E_\O$, and setting likewise ${\cal H}^{\ell}_\cO = P_\cO\cH^{\cL}$, 
 we obtain an object $(\euL_\cO, V_\cO,\boldsymbol{L} )$ in the category ${\bf Ghyst}$
upon defining the isometry $V_\cO : \cH^{\ell}_\cO \to \cH^\cL$ by $P_{\cO}\varphi \mapsto P_{\cO}\varphi$,
$\varphi \in \cH^\cL$, and $\ell_\cO = V_{\cO}^* {\cal L}V_\cO$, etc.   But
we would also like to encode that $\cO$ be globally hyperbolic. A very convenient set of globally
hyperbolic submanifolds of a globally hyperbolic spacetime is the set of diamond regions. Here, we say that
$\cO$ is a {\it diamond} if it is of the form $\cO = D(\O)$ where $\O$ is any open subset of some Cauchy-surface
$\Sigma$ in $\cM$; $D(\O)$ here denotes the (open) domain of dependence of $\O$, defined as the
open interior of all points $q$ in $\cM$ such that {\it every} inextendible causal curve passing through $q$
also meets $\O$, cf.\ Figure 1 (see \cite{Wald.GR} for discussion).
\begin{center}
\epsfig{file=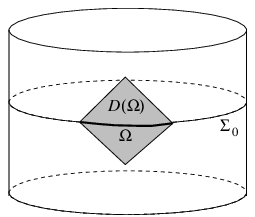,width=5.5cm}
\end{center}
{\footnotesize {\bf Figure 1.}  The figure depicts a diamond $D(\O)$ based on a segment $\O$ of the
Cauchy-surface $\Sigma_0$ in a spacetime of the form $\bR \times S^1$.}
\\[6pt]
Coming back to the algebraic construction of $\cO = D(\O)$, 
let the spacetime $\cM$ be represented as $\bR \times \Sigma$ according to the foliation into
Cauchy-surfaces induced by the given ghyst $\boldsymbol{L}$. Let us assume (without loss
of generality) that $\O$ is an
open subset of the Cauchy-surface $\Sigma_0 = \{0\} \times \Sigma$ corresponding to the value 0
of the foliation-time parameter. 
We now wish to characterize
``strips'' that entirely lie
in $D(\O)$, i.e.\ subsets of $D(\O)$ of the form $I \times \O^* \subset \bR \times \Sigma$,  where
$I$ is an open interval containing $0$ and $\O^*$ an open subset of $\O$ (cf.\ Fig.\ 2). In other words, we have to characterize
the corresponding subspaces $\cH^{\cL,\infty}_{I \times \O^*}$ of smooth
sections of the spinor bundle over $\cM$ with support in $I \times \O^* \subset D(\O)$. 
\begin{center}
\epsfig{file=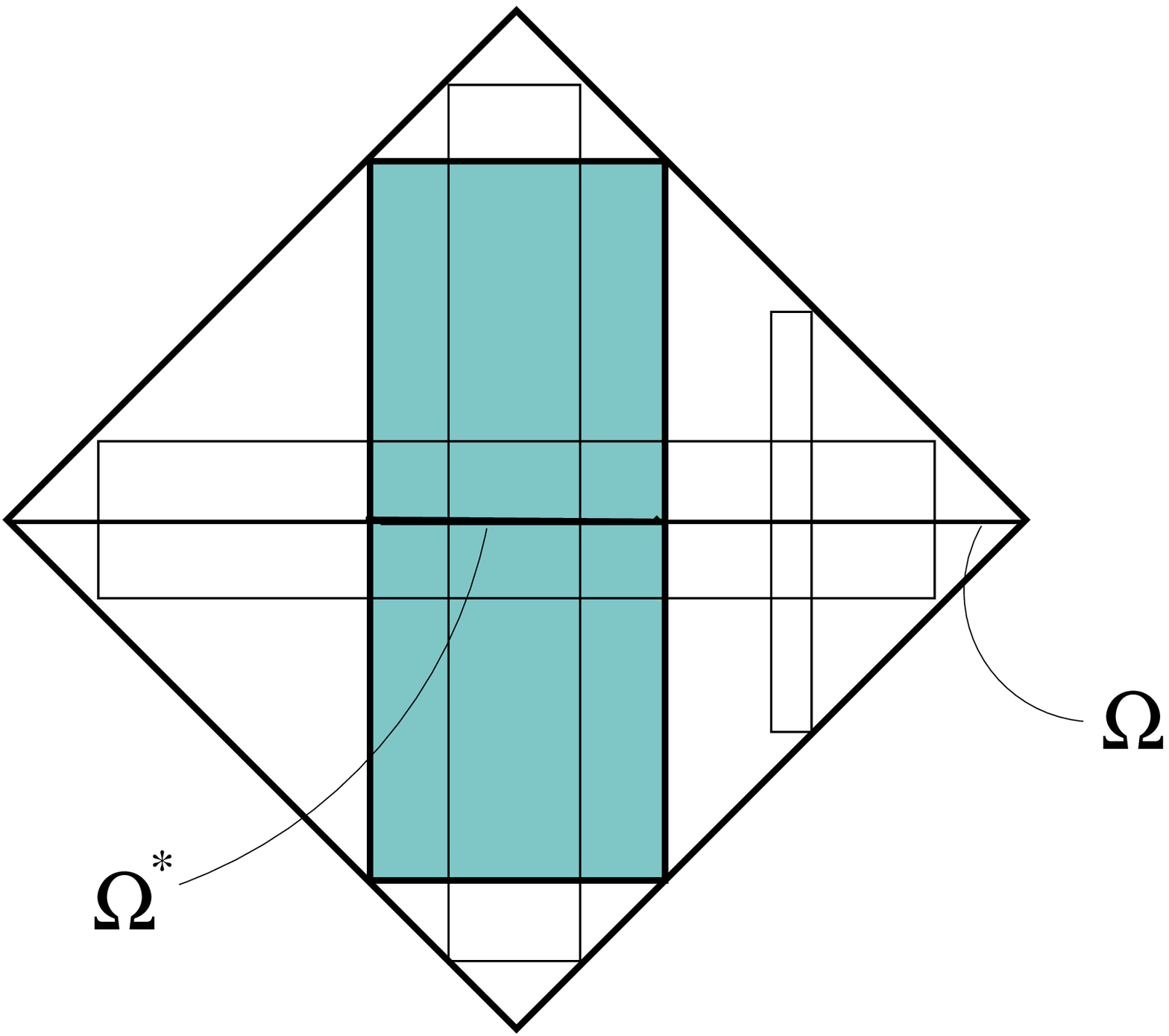,width=6cm}
\end{center}
{\footnotesize {\bf Figure 2.} The shaded region shows the strip $I \times \O^*$ with the time-interval $I$
chosen such that $I \times \O^*$ just remains inside $D(\O)$. Other strips with this property are also
indicated.}
\\[6pt]
To this end, let $\psi$ be any solution to the Dirac-equation
with Cauchy-data supported at foliation-time $t=0$ in the complement of $\O$,
i.e.\ $\psi$ has Cauchy-data supported in $\Sigma_0\backslash\O$.
Then supp$(\psi) \cap  D(\O) = \emptyset$, by the definition of $D(\O)$ and the causal propagation property of solutions
to the Dirac-equation. We thus define
 \begin{equation}
\cH^{\cL,\infty}_{[\Sigma_0\backslash\O]} := \{ \psi\in \cH^{\cL,\infty}\; : \; D^\cL\psi =0 \quad {\rm and}\quad  
\psi|_{\O} = 0  \}.  
\label{hil}
\end{equation}
Then, if $\varphi \in \cH^{\cL,\infty}$ has the
property to be supported on a strip $I \times \O^*$ remaining inside $D(\O)$,   it follows that
\begin{equation}
 \langle \psi,  X \varphi\rangle =0  \qquad \forall\ \ \ \psi \in  \cH^{\cL,\infty}_{[\Sigma_0\backslash \O]}\,, \ \
    X\in \{\cL,D^\cL,J^\cL,\gamma^\cL  \} \,.\label{glei}
\end{equation}
Conversely, if $\varphi$ is supported on $I \times \O^*$, then one can show that the property
\eqref{glei} implies that $\varphi$ must also be supported inside $D(\O)$, and this can now
be used to characterize that the strip $I \times \O^*$ lies inside $D(\O)$. Therefore, we define
$\cH^{\cL,\infty}_{I \times \O^*}$ as the set of all those $\varphi \in \cH^{\cL,\infty}$ which
are supported in  $I \times \O^*$, with $I$ chosen such that \eqref{glei} holds.
 
Using this definition, the subspace $\cH^{\ell}_{D(\O)} \equiv \cH^{\cL}_{D(\O)}$ is defined to be the Hilbert sub-space
of $\cH^\cL$ generated  by
all the  $\cH^{\cL,\infty}_{I \times \O^*}$, 
and that then will give rise to an orthogonal projector $P_{D(\O)} = V_{D(\O)}V_{D(\O)}^*$
onto $\cH^{\cL}_{D(\O)}$ with all the desired properties to the extent that $(\euL_{D(\O)},V_{D(\O)},\boldsymbol{L})$
really corresponds to an open, globally hyperbolic submanifold of the spacetime $\cM$ described by
$\boldsymbol{L}$.
In particular, the sub-category ${\bf Geo}$ will be defined as having as objects all the isometries
$V$ that can be constructed in this way.  The morphisms in {\bf Geo} will then be defined
as pairs of (partial) isometries constructed in a similar manner, such that the object class
remains invariant under the action of the morphisms.

We should finally remark that in this algebraic construction of $D(\O)$ only the algebraic and
Hilbert-space structure of the ghyst $\boldsymbol{L}$ was used. Hence, this construction   will also be well defined
in the noncommutative case: Suppose we are given a ghyst $\boldsymbol{L}$ with (noncommutative) 
spectral triples $(\cR_t , \cH_t,D_t,J,\gamma_t)$ for each $t$, and suppose we are given a
``sub-spectral triple'', at time $t=0$ say,  defined via some isometry (projection) $E$. Thus we have the analogue of
$\Sigma_0$ and $\O$ in the above construction, and we can use the definitions (\ref{hil}) and (\ref{glei})
to construct the corresponding diamonds. (The complement $\Sigma_0\backslash\O$ is of course 
given by the projection $1-E$.)

We should furthermore stress that for an arbitrary, given spectral triple it might
happen that there don't exist any nontrivial
isometries which meet all our requirements. Examples for ghysts 
with such time slices that do not possess any submanifolds shall be given in a forthcoming paper.
For these ghysts there  do not exist any globally hyperbolic submanifolds in the above  sense, or, in more physical terms,
it would not be possible to define any operations that are localized in this sense.
One might therefore ask whether (at least for such examples) there is a less restrictive, yet
 natural notion of submanifolds 
generalizing our present concept based on isometries,
 and whether such a notion could still be used to formulate restrictions on the QFT functor.
We hope to return to this problem after examination  of generally covariant quantum field theories over such ghysts.  
At any rate, we believe that a physically meaningful notion of submanifolds should be defined via the QFT functor,
i.e.\ in terms of the observables of a quantum theory that decribes the dynamical coupling of geometry to matter.

\subsection{The simplest noncommutative example and the notion of Morita equivalence}

The simplest example of a noncommutative ghyst is obtained by taking a commutative ghyst over some algebra
$\cL = C^\infty_0(\cM)$ and tensoring $\cL$ with some finite-dimensional matrix algebra, 
$$\cL_n := \cL \otimes M_n(\bC). $$
Tensoring the Hilbert space $\cH^L$ of the ghyst over $\cL$ with $\bC^n$ and the other data of the ghyst with the identity
on $\bC$ one obviously obtains a ghyst over $\cL_n$. It is, in fact, possible to slightly generalize this example as
there is the freedom to add a $su(n)$-gauge potential to the Dirac-operator. 
Evidently, there do exist exist many globally hyperbolic submanifolds in this case, as one may simply 
replace the isometry $V$ corresponding to such a submanifold of $\cM$ by $V\otimes {\rm id}_n$
to obtain a noncommutative one. In fact, as is easily shown, {\em all} globally hyperbolic submanifolds of
ghysts over $\cL_n$ are obtained this way.

It is also immediately clear that in the same spirit we can construct generally covariant quantum field theories over 
such a space:  Given the restriction of some QFT functor {\sf A}  to the category of globally hyperbolic submanifolds
of $\cM$, we can obtain a functor by setting e.g. in case of a Dirac-field
\[ \mbox{\sf A}_n = \mbox{\sf A}\otimes {\rm id}_n\,, \]
that is to say, by just treating the matrix degrees of freedom as mere multiplicities in the quantization procedure.
It is fairly obvious that all the functorial properties are fullfilled.
The same
functor would therefore also arise as a possible generally covariant QFT over $\cM$.
Vice versa, given any QFT-functor, {\sf B}$_n$ say, over $\cL_n$, one can construct a QFT-functor over $\cM$
by setting
\[ \mbox{\sf B}_n = {\rm id}_{n}\otimes_{M_n(\bC)} \mbox{\sf B}\,, \] 
where the tensor product over the algebra $M_n(\bC)$ of a right module (representation space) $V_1$ of $M_n(\bC)$
and  a left-$M_n(\bC)$-module $V_2$ is defined as $V_1\otimes_{M_n(\bC)} V_2 = V_1\otimes V_2/\sim$,
i.e.\ by dividing the usual tensor product (over $\bC$) by the equivalence relation
$$(\psi_1 m) \otimes \varphi_2 \sim  \psi_1 \otimes (m \varphi_2) \,,\qquad\qquad \forall\ m \in M_n(\bC) . $$
Note the isomorphism $\bC^n \otimes_{M_n(\bC)} \bC^n = \bC $ from which the above statement follows.
Thus, the sets of all generally covariant quantum field theories over $\cM$ respectively $\cL_n$ are equal.
The above construction is a special case of {\em Morita equivalence} (see, e.g., \cite{GB-F-V:book} and 
references cited there for further discussion):

Two ($C^*$)-algebras $\cA,\cB$ are called Morita equivalent if there are $\cA-\cB$-modules $W,W^*$, such that for
every $\cA$-module $V_\cA$ one can construct a $\cB$-module as $V_\cA\otimes_\cA W$, and similarly one can construct
$\cA$ modules from $\cB$-modules $V_\cB$ as $V_\cB\otimes_\cB W^*$. Thus, Morita equivalent algebras have the same 
representation theory and, as can easily be shown, also the same set of equivalence classes
of ghysts. In the light of the above example it is immediately clear that they will
also lead to the same set of QFT functors.

There do, of course, also exist many nontrivial examples of ghysts. In a forthcoming paper we shall 
show that one can construct the QFT functor corresponding to the free Dirac-field over the full subcategory 
{\bf Geo}, i.e.\ one can quantize  
the Dirac-field over all noncommutative spacetimes (in the above sense) simultaneously and in accordance with the principle of
general covariance.

\section{Outlook}
In this paper we have proposed a framework for generally covariant QFT over noncommutative spacetimes with a commuting 
time variable.

The restriction to a commuting time is certainly a severe drawback for at least two reasons:

First of all, it is at variance with (the physical idea of) the principle of general covariance, since it singles
out a preferred time coordinate.
Secondly, heuristic arguments as the one given in \cite{DFR} suggest that one should not expect that
time commutes with all the spatial variables in a full quantum theory describing the coupling of geometry to matter.

However, as our knowledge about spacetime is gained from scattering experiments, which seem to require at least an asymptotic
notion of ``large times'', our framework might well serve as an effective theory, which describes scattering
processes correctly, but which does not attempt to assign a geometrical (in the sense of ghysts) structure
to the ``interaction region''.
In this context, it should also be noted that our variable $t$ does -- similar to the coordinates $x^\mu$ for the one-particle
approximation to the free Dirac-field -- not have a direct physical meaning, but should only be viewed as an auxilary variable.

Nevertheless, it will be important to generalize the notion of ghysts to  cover also manifolds with a noncommuting 
time variable. Since  -- apart from one exception -- our axioms for ghysts only make essential use of some properties of the 
time evolution operators $U_{t,s}$ but not of the variable $t$ itself, we expect that such a generaliztion will be possible.

The mentioned exception is the algebra $\cL$ formed by the smooth sections in the bundle $\{\cR_t\}_{t\in\bR}$
over $\bR$. It is, of course, fairly obvious
how to modify the algebra $\cL$ in case of a noncommuting time, but that is not so for the family of algebras
$\cR_t\;$: Each algebra $\cR_\tau$ for fixed $\tau$ is supposed to be the projection of $\cL$ on the eigenspace
$\cH_\tau^\cL$ of $t$ with eigenvalue $\tau$. But then, if $t$ is not in the center of $\cL$, $\cR_\tau$ will in general 
fail to be an algebra. However, as in our definition we need a spectral triple for each $\cR_\tau$ 
the existence of these algebras is indispensable.

On the other hand, one might anyway ask whether the algebra $\cL$ should really be included in the data
that describe geometry:

From the physical point of view, it is the only part of a ghyst that is not directly accessible in scattering experiments.
Viewed from a mathematical perspective, as A. Connes has pointed out, it is possible that -- given the remaining data --
$\cL$ can be reconstructed as the (largest) algebra fulfilling the axioms of ghysts (resp.\ spectral triples) with those data.   
In fact, in \cite{CL} this reconstruction of the algebra has been used successfully to construct new examples of 
spectral triples. 

If such a program can be carried through, then it would enable us to define generally covariant QFT directly in terms of
the eigenvalues of the Laplacian -- or, to be more precise, the full set of scattering data. Recalling the introduction 
of this paper, one would thus try to ``hear'' the geometry of spacetime.

Of particular interest in this context is the observation by Schroer \cite{Schroer,SchWie} that one can
construct families of certain algebras of quantum field observables (``wedge algebras''), 
whose Tomita-Takesaki modular objects with
respect to the vacuum state have geometrical significance, from a factorizing scattering matrix in
1+1 spacetime dimensions. This provides a link between the ``form-factor program'', which aims at 
reconstructing quantum field theories from their scattering data, and the operator algebraic approach
to quantum field theory. First results in this direction have been obtained recently \cite{Lechner,BuLech}.
One might therefore also envisage the possibility that the geometry of spacetime can be reconstructed from
a form of geometric meaning associated with the Tomita-Takesaki modular objects corresponding to certain
preferred algebras and states of quantum matter observables, cf.\ the 
``geometric modular action principle'' in quantum field theory \cite{BMS,BDFS}. The elements of
${\cal L}$ may then be thought of as being coupled to, and even constructed out of, the elements
of these preferred algebras. Moreover, extrapolating on Schroer's ideas, 
these preferred algebras and states should be reconstructable from scattering data. Speculative as this
perspective may still appear, it would open a fascinating connection between the geometrical concepts
of Connes' spectral geometry and the geometrical contents of Tomita-Takesaki modular objects in terms
of the geometric modular action principle in quantum field theory. It is tempting to see here also
a connection to ideas on holography in the context of algebraic quantum field theory in the setting proposed 
by Rehren
\cite{Rehren1,Rehren2}.

The description of geometry in terms of scattering data, and in particular the spectrum of the Hamiltonian 
(resp., the Dirac-operator) is also essential for the construction of the (euclidean) ``noncommutative gravitational action'',
i.e.\ Connes' spectral action.  It would certainly be beneficial for the program outlined here to
work out a Lorentzian analogue thereof, resp.\ of the field equation derived from such an action. 
However, it is  not clear whether this step is really indispensable to achieve our ultimate aim, namely
to find the constraints on the QFT functor {\sf A} that ensures the dynamical coupling of matter 
to geometry. We should mention at this point the nice toy model presented in \cite{Rov},
where (a variant of) the spectral action over a particularly simple noncommutative Riemannian manifold has been quantized,
without preserving the ``diffeomorphism'' invariance, however.
Nevertheless, one obtains in this zero-dimensional example  a very clear and appealing picture of the
algebra of observables of such models, which might turn out 
helpful for the construction of a generally covariant QFT theory over such spaces, and in particular 
of  $(0+1)$-dimensional quantum gravity.    

At any rate, it is  the next principle task for our program to find  additional requirements on {\sf A} 
which enforce certain physically desired properties.
The appropriate restrictions to ensure locality and the existence of a causal dynamical law will be given
in a forthcoming paper; other issues like e.g.\ CPT invariance are under investigation.
We believe that (physically) appropriate constraints on {\sf A} will also
turn out to be constraints on the ``source category'' of {\sf A}.
In particular, one might hope that the constraints which correspond to the
dynamical coupling of geometry to matter will severely restrain the category {\bf Geo}, thereby maybe even 
exluding all classical (commutative)  manifolds, thus
leading to a refined picture of  spacetime at small scales.

\end{document}